\documentclass[twocolumn, preprint]{aastex63}
\usepackage{amsmath,amstext}
\usepackage[T1]{fontenc}
\usepackage{apjfonts} 
\usepackage{natbib}
\usepackage{microtype}
\usepackage{longtable}
\usepackage{verbatim}
\usepackage{upgreek}
\usepackage{multirow}
\usepackage{hyperref}
\received{\today}
\revised{XXX}
\accepted{XXX}
\submitjournal{ApJL}

\shorttitle{Spatially Resolved Dust Attenuation with {\it JWST} NIRISS}
\shortauthors{Matharu et al.}
\graphicspath{{./}{figures/}}

\begin{document}

\title{A First Look at Spatially Resolved Balmer Decrements at $1.0<z<2.4$ from {\it JWST} NIRISS Slitless Spectroscopy}

\correspondingauthor{Jasleen Matharu}
\email{jasleen.matharu@nbi.ku.dk}

\author[0000-0002-7547-3385]{Jasleen Matharu}
\affiliation{Cosmic Dawn Center (DAWN), Denmark}
\affiliation{Niels Bohr Institute, University of Copenhagen, Jagtvej 128, DK-2200 Copenhagen N, Denmark}

\author[0000-0002-9330-9108]{Adam Muzzin}
\affiliation{Department of Physics and Astronomy, York University, 4700 Keele Street, Toronto, ON, M3J 1P3, Canada\\}

\author[0000-0001-8830-2166]{Ghassan Sarrouh}
\affiliation{Department of Physics and Astronomy, York University, 4700 Keele Street, Toronto, ON, M3J 1P3, Canada\\}

\author[0000-0003-2680-005X]{Gabriel Brammer}
\affiliation{Cosmic Dawn Center (DAWN), Denmark}
\affiliation{Niels Bohr Institute, University of Copenhagen, Jagtvej 128, DK-2200 Copenhagen N, Denmark}

\author[0000-0002-4542-921X]{Roberto Abraham}
\affiliation{Dunlap Institute for Astronomy and Astrophysics, 50 St.~George Street, Toronto, Ontario, M5S 3H4, Canada\\}

\author[0000-0003-3983-5438]{Yoshihisa Asada}
\affiliation{Department of Astronomy and Physics and Institute for Computational Astrophysics, Saint Mary's University, 923 Robie Street, Halifax, Nova Scotia B3H 3C3, Canada\\}
\affiliation{Department of Astronomy, Kyoto University, Sakyo-ku, Kyoto 606-8502, Japan\\}

\author[0000-0001-5984-0395]{Maru{\v s}a Brada{\v c}}
\affiliation{University of Ljubljana, Department of Mathematics and Physics, Jadranska ulica 19, SI-1000 Ljubljana, Slovenia}
\affiliation{Department of Physics and Astronomy, University of California Davis, 1 Shields Avenue, Davis, CA 95616, USA}

\author[0000-0001-8325-1742]{Guillaume Desprez}
\affiliation{Department of Astronomy and Physics and Institute for Computational Astrophysics, Saint Mary's University, 923 Robie Street, Halifax, Nova Scotia B3H 3C3, Canada\\}

\author[0000-0003-3243-9969]{Nicholas Martis}
\affiliation{Department of Astronomy and Physics and Institute for Computational Astrophysics, Saint Mary's University, 923 Robie Street, Halifax, Nova Scotia B3H 3C3, Canada\\}
\affiliation{National Research Council of Canada, Herzberg Astronomy \& Astrophysics Research Centre, 5071 West Saanich Road, Victoria, BC, V9E 2E7, Canada\\}

\author[0000-0002-8530-9765]{Lamiya Mowla}
\affiliation{Dunlap Institute for Astronomy and Astrophysics, 50 St.~George Street, Toronto, Ontario, M5S 3H4, Canada\\}

\author{Ga\"el Noirot}
\affiliation{Department of Astronomy and Physics and Institute for Computational Astrophysics, Saint Mary's University, 923 Robie Street, Halifax, Nova Scotia B3H 3C3, Canada\\}

\author[0000-0002-7712-7857]{Marcin Sawicki}
\affiliation{Department of Astronomy and Physics and Institute for Computational Astrophysics, Saint Mary's University, 923 Robie Street, Halifax, Nova Scotia B3H 3C3, Canada\\}

\author[0000-0002-6338-7295]{Victoria Strait}
\affiliation{Cosmic Dawn Center (DAWN), Denmark}
\affiliation{Niels Bohr Institute, University of Copenhagen, Jagtvej 128, DK-2200 Copenhagen N, Denmark}

\author[0000-0002-4201-7367]{Chris J. Willott}
\affiliation{National Research Council of Canada, Herzberg Astronomy \& Astrophysics Research Centre, 5071 West Saanich Road, Victoria, BC, V9E 2E7, Canada\\}

\author[0000-0003-4196-5960]{Katriona M.L. Gould}
\affiliation{Cosmic Dawn Center (DAWN), Denmark}
\affiliation{Niels Bohr Institute, University of Copenhagen, Jagtvej 128, DK-2200 Copenhagen N, Denmark}

\author[0009-0006-4881-3299]{Tess Grindlay}
\affiliation{National Research Council of Canada, Herzberg Astronomy \& Astrophysics Research Centre, 5071 West Saanich Road, Victoria, BC, V9E 2E7, Canada\\}
\affiliation{Department of Physics and Astronomy, University of Victoria, Victoria, BC, V8P 5C2, Canada}

\author[0000-0001-9414-6382]{Anishya T. Harshan}
\affiliation{Department of Mathematics and Physics, Jadranska ulica 19, SI-1000 Ljubljana, Slovenia\\}


\def\editR{\textbf}



\def\editR{\textbf}
\def\ha{H$\upalpha$}
\def\hb{H$\upbeta$}

\begin{abstract}

We present the first results on the spatial distribution of dust attenuation at $1.0<z<2.4$ traced by the Balmer Decrement, \ha/\hb, in emission-line galaxies using deep {\it JWST} NIRISS slitless spectroscopy from the CAnadian NIRISS Unbiased Cluster Survey (CANUCS). \ha~and \hb~emission line maps of emission-line galaxies are extracted and stacked in bins of stellar mass for two grism redshift bins, $1.0<z_{grism}<1.7$ and $1.7<z_{grism}<2.4$. Surface brightness profiles for the Balmer Decrement are measured and radial profiles of the dust attenuation towards~\ha~, $A_{\mathrm{H}\alpha}$, are derived. In both redshift bins, the integrated Balmer Decrement increases with stellar mass. Lower mass ($7.6\leqslant$Log($M_{*}$/M$_{\odot}$)$<10.0$) galaxies have centrally concentrated, negative dust attenuation profiles whereas higher mass galaxies ($10.0\leqslant$Log($M_{*}$/M$_{\odot}$)$<11.1$) have flat dust attenuation profiles. The total dust obscuration is mild, with on average $0.07\pm0.07$ and $0.14\pm0.07$ mag in the low and high redshift bins respectively. We model the typical light profiles of star-forming galaxies at these redshifts and stellar masses with \texttt{GALFIT} and apply both uniform and radially varying dust attenuation corrections based on our integrated Balmer Decrements and radial dust attenuation profiles. If these galaxies were observed with typical {\it JWST} NIRSpec slit spectroscopy ($0.2\times0.5^{\prime\prime}$ shutters), on average, \ha~star formation rates (SFRs) measured after slit-loss corrections assuming uniform dust attenuation will overestimate the total SFR by $6\pm21 \%$ and $26\pm9 \%$ at $1.0\leqslant z < 1.7$ and $1.7\leqslant z < 2.4$ respectively.



\end{abstract}



\section{Introduction}
\label{sec:intro}

It is now well-established that galaxy evolution is a complicated interplay between various physical processes, including the inflow and outflow of gas, as well as its recycling. Star formation in a galaxy primarily depends upon its gas reservoir \citep{Schmidt1959,Kennicutt1998,Kennicutt2012}. How this star formation propagates through galaxies can determine how their structural growth proceeds.

Determining where the star formation process starts and finishes in a galaxy is only possible with spatially resolved studies of star formation tracers. At $z\gtrsim 0.5$, ground-based integral field spectroscopy and space-based slitless (or ``grism") spectroscopy with {\it Hubble Space Telescope}'s (HST) {\it Wide Field Camera 3} (WFC3) 
have allowed us to measure the spatial distribution of ongoing star formation in galaxies using the \ha~emission line as a star formation tracer. Some of these studies have revealed the inside-out growth and subsequent inside-out quenching of galaxies \citep{ForsterSchreiber2009,Schreiber2018a,Law2009,Nelson2012,Nelson2013a,Nelson2015,Tacchella2015a,Wisnioski2015,Wisnioski2019,Wilman2020,Matharu2021a,Matharu2022, Noirot2022b}.

Whilst \ha~emission predominantly traces emission from young O and B stars, it is susceptible to dust attenuation. Since dust preferentially attenuates light at bluer wavelengths, dust attenuation towards \ha~emitting regions can be measured by comparing the ratio of the \ha~flux to the \hb~flux. Known as the Balmer Decrement, the \ha/\hb~ratio has been shown to increase with SFR, stellar mass and attenuation towards the stars \citep{Calzetti1999,Wild2011,Dominguez2012,Momcheva2013,Price2014,Reddy2015,Shivaei2015} but not vary significantly with redshift out to $z\sim6$ \citep{Shapley2022,Shapley2023}. Spatially resolved measurements of the Balmer Decrement at high redshift will help complete the picture of galaxy assembly, revealing the location of obscured star formation in galaxies.

The power of space-based slitless spectroscopy lies in its high spatial resolution, providing 2D emission-line maps of high-redshift galaxies. However, the lack of a slit mask leads to 2D spectra for all the sources in the field-of-view. The result is contamination of spectra from neighbouring sources and high backgrounds. These issues limit the sensitivity of space-based grism spectroscopy. Conversely, multi-object slit spectroscopy can reach higher sensitivities but has limited spatial resolution for a limited sample selected from pre-imaging. To account for the lack of spatial coverage, slit loss corrections need to be applied. With {\it JWST} NIRSpec, this has been done by using the NIRCam photometry of the galaxies, overlaying the shutter position and comparing how much flux passes through the shutter versus the total flux measurement from the photometry \citep{Larson2023}. The spectra are then scaled to match the total flux from the photometry. The Balmer Decrement is then measured from total line fluxes in the spectra, providing a single value of $A_{\mathrm{H}\alpha}$ that is used to dust-correct SFR measurements. This dust correction assumes light from the galaxy that is outside of the slit is attenuated by the same amount as the light from the galaxy that passes through the slit. However, spatially resolved Balmer Decrements measured from {\it HST} WFC3 space-based slitless spectroscopy have shown that $9.0\leqslant$Log($M_{*}$/M$_{\odot}$)$<11.0$ galaxies at $z\sim1.4$ have negative dust attenuation profiles that are concentrated within the inner 1 kpc \citep{Nelson2016a}, indicating that this assumption may be invalid. In an era when many studies will exploit {\it JWST} NIRSpec to study SFRs and galaxy assembly through cosmic time, it is timely to test the validity of such assumptions.

In this paper, we extend high-redshift spatially resolved Balmer Decrement measurements out to $z\sim2.4$, into the peak of cosmic star formation \citep{Madau&Dickinson2014}, with spatially resolved \ha~and~\hb~emission line maps from {\it JWST} NIRISS slitless spectroscopy.

All magnitudes quoted are in the AB system, logarithms are in base 10 and we assume a $\Lambda$CDM cosmology with $\Omega_{m}=0.307$, $\Omega_{\Lambda}=0.693$ and $H_{0}=67.7$~kms$^{-1}$~Mpc$^{-1}$ \citep{Planck2015}.

\section{Dataset}
\label{sec:methodology}

\begin{figure*}
	\centering\includegraphics[width=\textwidth]{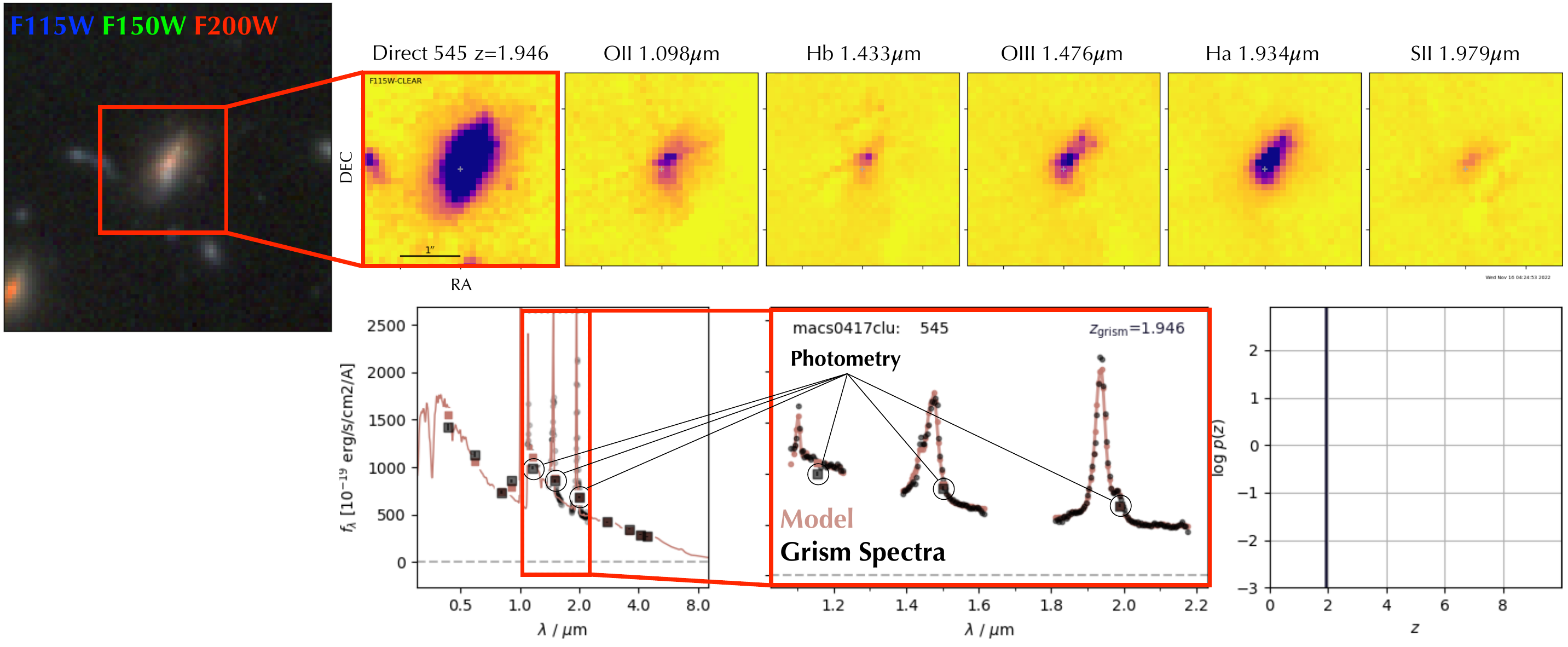}
    \caption{Visualisation of example \texttt{Grizli} data products for a galaxy in the MACS J0417-1154 cluster field that is in our sample. Top row: RGB thumbnail, blotted F115W thumbnail and drizzled emission line maps. Titles include the observed wavelength of each emission line. Bottom row: Redshift fit with photometry. Small circular black points are from the grism spectra and large black squares are photometric points. The SED model is shown in red. The Log p($z$) plot is shown to the right, with a well-determined grism redshift of 1.946 and a stellar mass of Log($M_{*}$/M$_{\odot}$)$=10$.}
    \label{fig:data_viz}
\end{figure*}

We use {\it JWST} observations of galaxies in the background of the $z=0.4$ galaxy cluster MACS J0417-1154 that were taken from 12th-17th October 2022 as part of the NIRISS GTO Program \#1208, The Canadian NIRISS Unbiased Cluster Survey (CANUCS, \citealt{Willott2022}). Observations consist of two NIRISS pointings, one centered on the cluster center and the other coincident with a flanking field\footnote{See the pointing layouts of both fields at \url{canucs-jwst.com}.} that has existing {\it HST} WFC3/UVIS imaging (HST-G0-16667, PI: Brada\v{c}). Each field is observed with both the GR150R and GR150C grisms through the F115W, F150W and F200W filters. Total exposure times are 10822 seconds in F115W, 5411 seconds in F150W and F200W for the flanking field and 19240 seconds in all three filters for the cluster field. We process all the imaging and slitless spectroscopy with the Grism redshift \& line analysis software, \texttt{Grizli} \citep{Grizli2022}. \texttt{Grizli} performs full end-to-end processing of space-based slitless spectroscopic datasets. For full details on \texttt{Grizli} and its data products, we refer the reader to \cite{Matharu2021a}, \cite{Simons2020a} and \cite{Noirot2022}. In summary, raw data is downloaded from the Mikulski Archive for Space Telescopes (MAST) and pre-processed for cosmic rays, flat-fielding, sky subtraction, astrometric corrections and alignment \citep{Gonzaga2012,Brammer2015,Brammer2016}. Contamination models (which correct for overlapping spectra from nearby sources) for each pointing are generated and subtracted for each grism spectrum of interest. 

\subsection{Grism redshift fitting, Sample Selection \& Stellar Masses}
\label{sec:fitting}

When deriving grism redshifts for each source, we fit the grism spectra and multiwavelength photometry simultaneously. Figure~\ref{fig:data_viz} shows this fitting along with the resulting emission-line maps that are extracted from the data for a high signal-to-noise ratio (SNR) galaxy in the cluster field that is in our sample. For more details on the photometry and image processing see \cite{Noirot2022} and \cite{Asada2022}.

Fits for all sources are checked by eye for secure grism redshifts. We define a secure grism redshift by (1) The Spectral Energy Distribution (SED) model is a good fit to the photometric and grism spectroscopic points (red line in Figure~\ref{fig:data_viz}) and (2) there are at least two features that constrain the log p($z$). That is either two emission lines or one emission line and a Balmer break. This last criterion is the same as quality 1 for the SMACS grism redshift catalog \citep{Noirot2022}.

We then re-fit all sources with secure grism redshifts without photometry, but constrain the fitting to a narrow redshift range around the initially determined grism redshift. This often allows for cleaner emission line maps, especially in cases where the \texttt{Grizli} scaling to photometry leads to over- or under-subtraction of the continuum. Those galaxies with integrated Balmer Decrements from \texttt{Grizli} (see Section~\ref{sec:int_balmer_decs}) with errors larger than 100\% of their measured value are removed from the sample. Galaxies in the cluster field with lensing magnification greater than four are removed. The median magnification for galaxies in the sample that reside in the cluster field is two. The model of the cluster's lensing magnification distribution is obtained using \texttt{Lenstool} \citep{Kneib1993,Jullo2007}, further details of which are described in \cite{Strait2023} and Desprez 2023, in prep.

Measuring Balmer Decrement profiles with our dataset requires \ha~and~\hb~to fall within filters F115W and F150W or F150W and F200W. This sets a maximum redshift range of $1.0<z<2.4$ for our study. Our final sample consists of 56 galaxies in the cluster field and 61 galaxies in the flanking field that have both a \ha~and \hb~emission-line map extracted. During the grism redshift fitting process, we avoid biasing the sample towards dust-free galaxies by not imposing any SNR thresholds on the detections of \ha~and~\hb. Bad pixels and neighbouring sources in all the \ha~and \hb~emission line maps are then masked. We split this sample into two grism redshifts bins, $1.0<z_{grism}<1.7$ and $1.7<z_{grism}<2.4$, containing 58 and 59 galaxies respectively. We rescale our \ha~fluxes and surface brightness profiles down for the contribution of [\ion{N}{2}] using the \cite{Zahid2014} relation as in \cite{Nelson2016a} throughout this paper.

We calculate the stellar masses of our final sample of galaxies with \texttt{Dense Basis} \citep{Iyer2019} at the grism redshifts obtained from \texttt{Grizli} ($\pm$0.01). The \texttt{Dense Basis} atlas was generated with a flat specific star formation rate prior, \cite{Calzetti1999} dust law, \cite{Chabrier2003} Initial Mass Function (IMF), and an exponential reddening prior with scale values $0 \leqslant A_{v} \leqslant 4$. Stellar masses for galaxies in the cluster field are corrected for lensing magnification.

\subsection{Star Formation Main Sequence}
\label{sec:sfms}

\begin{figure}
	\centering\includegraphics[width=\columnwidth]{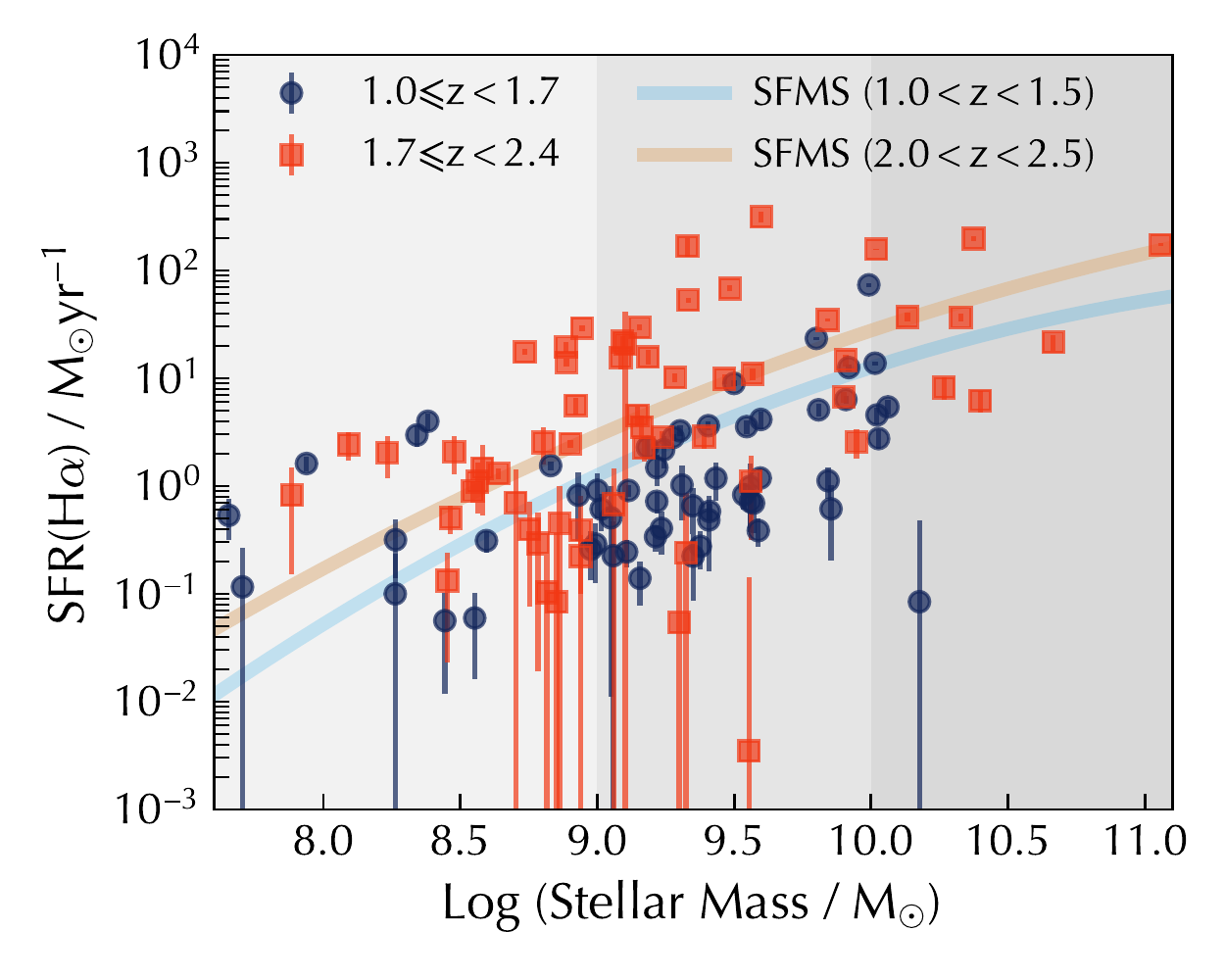}
    \caption{The star formation main sequence (SFMS) of our sample with dust-corrected SFRs using our Balmer Decrement measurements (Section~\ref{sec:int_balmer_decs}). Grey shaded regions delineate the stellar mass bins for the stacks. The UV+IR SFMSs within which the median redshift of our two samples fall from \cite{Whitaker2014a} are shown as the thick solid lines.}
    \label{fig:sfms}
\end{figure}

To place our sample of emission-line galaxies in context, we show its star formation main sequence (SFMS) in Figure~\ref{fig:sfms} with the ultraviolet (UV) + infrared (IR) SFMSs from \cite{Whitaker2014a} that overlap with the median redshifts of our two samples. All \ha~fluxes are dust-corrected using our integrated Balmer Decrement measurements on the individual galaxies (see Section~\ref{sec:int_balmer_decs}). \ha~fluxes for galaxies in the cluster sample are corrected for magnification. We calculate SFRs using the \ha~flux conversion from \cite{Kennicutt1998}, adapting it from a Salpeter IMF to a \cite{Chabrier2003} IMF using the method in \cite{Muzzin2010}. Both the high and low redshift samples follow the SFMSs from \cite{Whitaker2014a} well, with the low and high redshift samples having median offsets of $-0.4\pm0.1$ and $0.0\pm0.1$~dex from the \cite{Whitaker2014a} SFMSs respectively. Note that at Log($M_{*}$/M$_{\odot}$)$\lesssim 8.5$, we may be preferentially detecting galaxies with higher specific star formation rates (sSFRs).

\subsection{Integrated Balmer Decrements}
\label{sec:int_balmer_decs}

\begin{figure}
	\centering\includegraphics[width=\columnwidth]{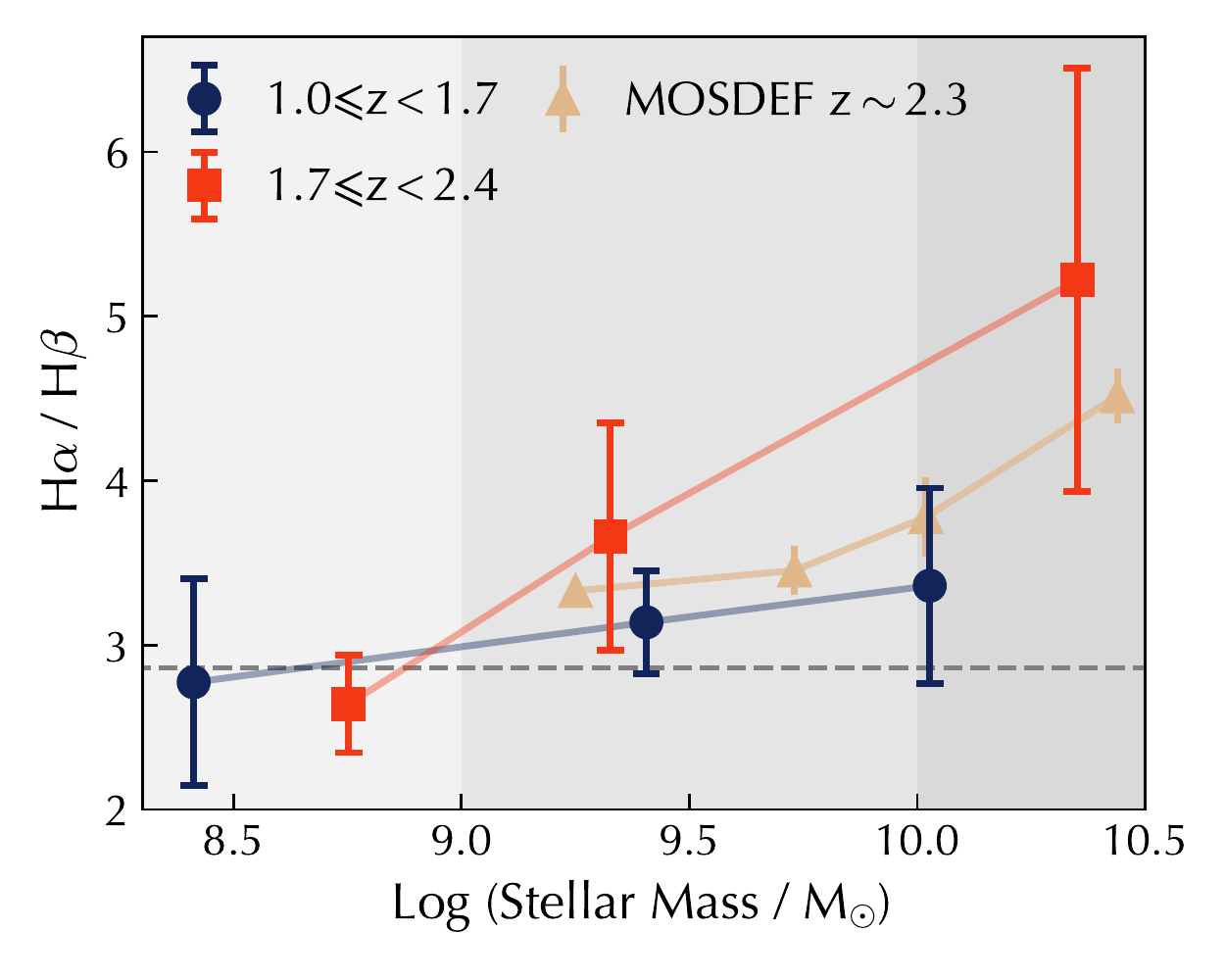}
    \caption{Integrated Balmer Decrements as a function of stellar mass. Grey regions show the different stellar mass bins. Grey dashed horizontal line shows the expected Balmer Decrement in the absence of dust attenuation (see Section~\ref{sec:balmer_decs}). Points are medians and plotted at the median stellar mass of each bin. Brown triangles show the results from MOSDEF \citep{Shapley2022}. All errors are errors on the median.}
    \label{fig:int_BDs}
\end{figure}

In Figure~\ref{fig:int_BDs}, we show our integrated Balmer Decrements as a function of stellar mass calculated using the integrated \ha\footnote{The \ha~flux is scaled down for the contribution of [\ion{N}{2}] as described in Section~\ref{sec:fitting}.}~and~\hb~fluxes output by \texttt{Grizli} and compare them to those measured using the MOSFIRE multi-object slit spectrograph as part of the MOSDEF survey \citep{Shapley2022}. Despite differences in the instrument, type of spectroscopy and extraction methods, our high and low redshift measurements are consistent with those from MOSDEF within the $1\upsigma$ uncertainties of our measurements.

\section{Analysis}

\subsection{Spatially Resolved Balmer Decrements}
\label{sec:balmer_decs}

Figure~\ref{fig:data_viz} represents what our data products can look like for galaxies with \ha~and~\hb~at the highest SNRs (\ha~SNR=363, \hb~SNR=73). However, the median \ha~and~\hb~SNRs for our sample are 22 and 7 respectively. Visual inspection of the \ha~and~\hb~maps in Figure~\ref{fig:data_viz} however demonstrate that dust can be clumpy and off-center. In this analysis we decide to stack our galaxies both to maximize SNR and average over all possible geometries for \ha~and~\hb~spatial distributions. Within each of our redshift bins, we stack the \ha~and~\hb~emission line maps of our galaxies in three bins of stellar mass: $7.6\leqslant$Log($M_{*}$/M$_{\odot}$)$<9.0$, $9.0\leqslant$Log($M_{*}$/M$_{\odot}$)$<10.0$ and $10.0\leqslant$Log($M_{*}$/M$_{\odot}$)$<11.1$. We use the inverse variance maps generated for each emission line map by \texttt{Grizli} to weight each pixel. Then we add a further weighting by the total flux density of the galaxy in the corresponding direct imaging filter within which each emission line is detected. This second weighting ensures no single bright galaxy dominates the final stack \citep{Nelson2015}. The \ha~and~\hb~maps for each stack are then added together and corrected for exposure time by dividing by the corresponding summed weight stack. For each stack, variance maps are $\sigma_{ij}^2 = 1/\sum{w_{ij}}$, where $w_{ij}$ is the weight map for each galaxy emission line map in the stack. In our low redshift bin, \hb~and [\ion{O}{3}] are close in wavelength such that [\ion{O}{3}] is not fully masked. We mask remaining [\ion{O}{3}] emission by hand. 

\begin{figure}
	\centering\includegraphics[width=\columnwidth]{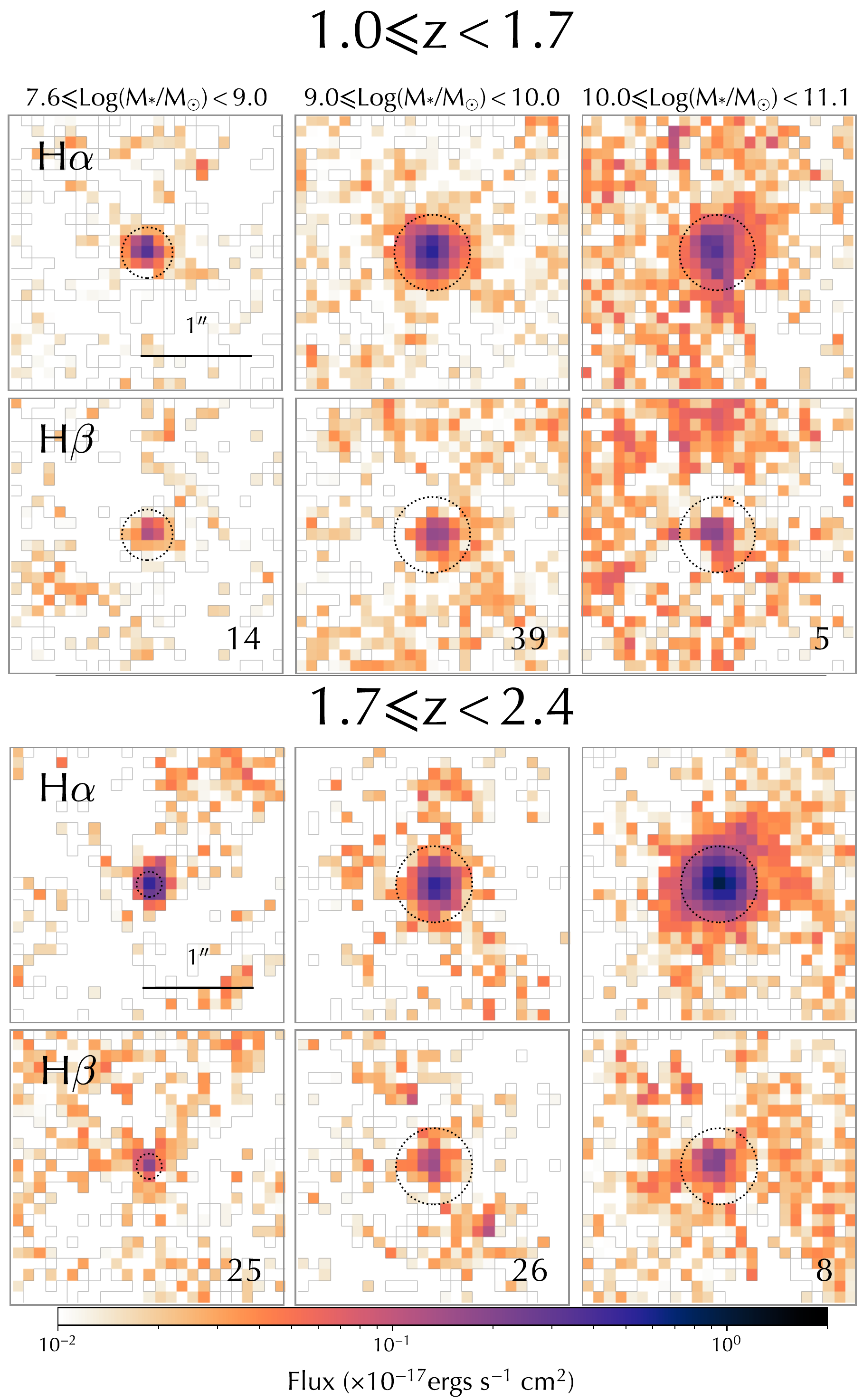}
    \caption{\ha~and \hb~stacks in bins of redshift and stellar mass. We measure surface brightness profiles out to the dotted circles, the largest of which has a radius of 3 kpc at the median redshift of the stack. The number of galaxies in each \ha~and~\hb~stellar mass stack is shown in the bottom right corner of each \hb~stack.}
    \label{fig:stacks}
\end{figure}

\begin{figure*}
	\centering\includegraphics[width=\textwidth]{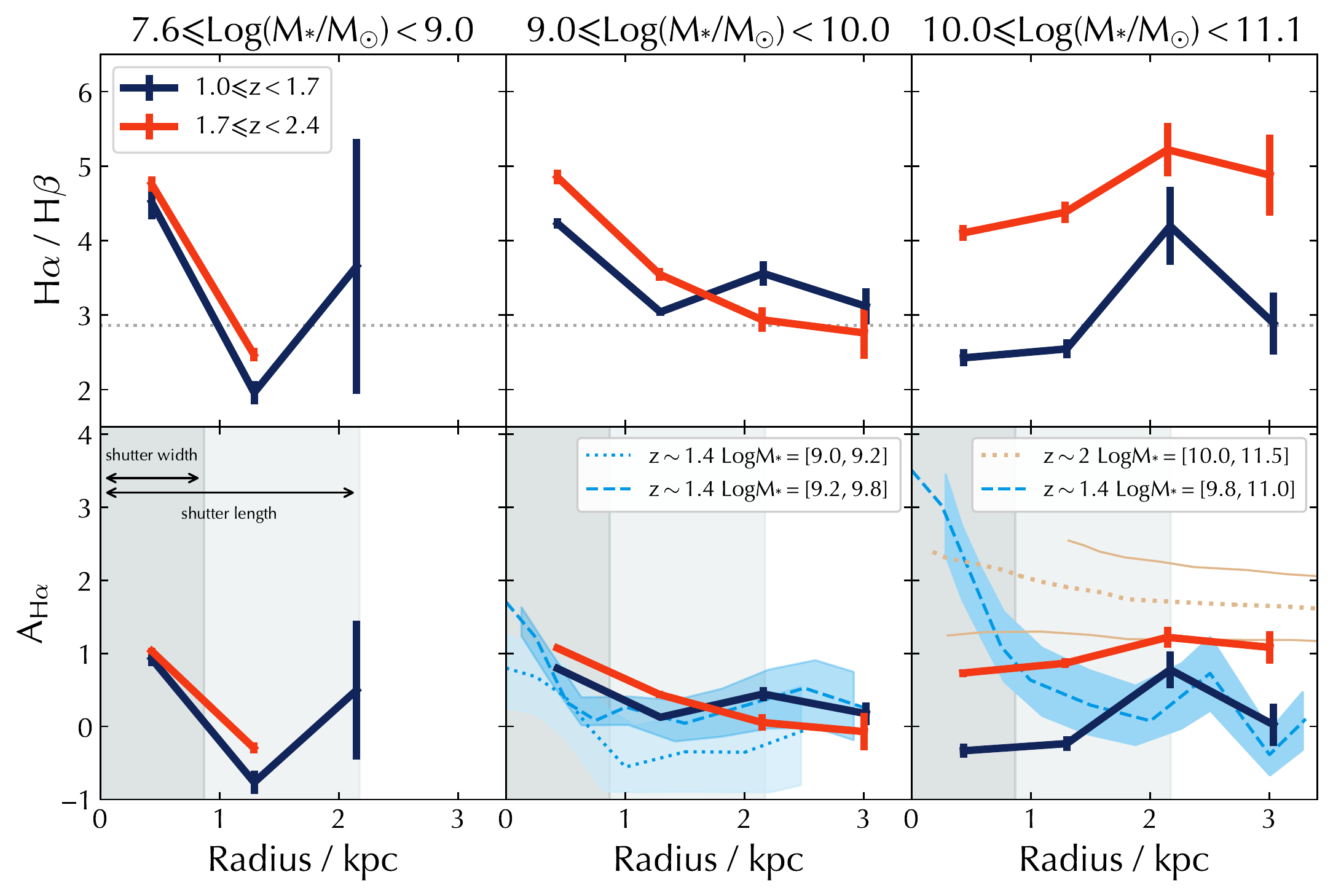}
    \caption{Radial profiles of the Balmer Decrement (top row) and of the dust attenuation towards \ha~emission assuming a \cite{Calzetti1999} dust law (bottom row). Blue and orange lines with error bars show our PSF-corrected measurements. The grey dotted horizontal lines show the expected Balmer Decrement in the absence of dust attenuation (see Section~\ref{sec:balmer_decs}). Dark and light grey shading delineate the width ($0.2^{\prime\prime}$) and length ($0.5^{\prime\prime}$) of a {\it JWST} NIRSpec shutter in kpc at $z=1.7$. Light blue dashed and dotted lines with error envelopes show the 3D-HST results from \cite{Nelson2016a}. The brown dotted line with solid lines delineating the $1\upsigma$ scatter is the result from \cite{Tacchella2018}.}
    \label{fig:BDec}
\end{figure*}

We then remove the effect of the point spread function (PSF) from our stacks following a method that is similar to that of \cite{Nelson2016a} but updated and modified for {\it JWST}. \texttt{Grizli} is used in combination with \texttt{WebbPSF} \citep{Perrin2012, Perrin2014} to create empirical monochromatic PSFs at the observed wavelengths of \ha~and \hb~for each galaxy in each stack. These are created for multiple positions across the NIRISS detector frame. A PSF at or near the location of the galaxy on the detector is then drizzled with the same parameters used for the emission line maps. The PSFs are then stacked, leading to unique PSFs for each \ha~and \hb~stack.  \texttt{GALFIT} \citep{Peng2002b, Peng2010a} is then used to fit each stack with a S\'ersic model and then create a PSF-unconvolved model based on parameters from the fits \citep{Szomoru2013}.

Figure~\ref{fig:stacks} shows the original PSF-convolved stacks zoomed in to a 25 x 25 pixel square region (original emission line map thumbnails are 80 x 80 pixels) with a pixel scale of $0.1^{\prime\prime}$. We measure the surface brightness profiles of the PSF-unconvolved \ha~and~\hb~stacks in ring apertures out to a maxmimum of 3 kpc (largest dotted circles in Figure~\ref{fig:stacks}) using the Python package \texttt{MAGPIE}\footnote{https://github.com/knaidoo29/magpie/}. \texttt{MAGPIE} accounts for the surface area of each pixel included within each ring aperture and allows the incorporation of our $\sigma$ images to calculate the errors for our surface brightness profiles. We then divide our \ha~surface brightness profiles by their corresponding \hb~surface brightness profiles to calculate radial Balmer Decrement profiles shown in the top row of Figure~\ref{fig:BDec}. The dotted horizontal lines show the expected \ha/\hb~ratio of 2.86 for Case B recombination and $T=10^{4}$K \citep{OsterbrockFerland2006}.

\subsection{Radial Gradients in Dust Attenuation}
\label{sec:A_ha}

We derive radial gradients in dust attenuation towards \ha~by first calculating the Balmer color excess, 
\begin{equation}
    E(\mathrm{H}\beta - \mathrm{H}\alpha) = 2.5~\mathrm{log}\left( \frac{(\mathrm{H}\alpha/\mathrm{H}\beta)}{2.86} \right),
\end{equation}
where (H$\alpha$/H$\beta$) is the measured Balmer Decrement. The attenuation towards \ha~is then:
\begin{equation}
    A_{\mathrm{H}\alpha}=\frac{E(\mathrm{H}\beta - \mathrm{H}\alpha)}{k(\lambda_{\mathrm{H}\beta})-k(\lambda_{\mathrm{H}\alpha})} \times k(\lambda_{\mathrm{H}\alpha}),
\end{equation}
where $k(\lambda_{\mathrm{H}\alpha})$ and $k(\lambda_{\mathrm{H}\beta})$ are the values of the adopted reddening curve at the wavelengths of the \ha~and~\hb~emission lines. We show radial profiles of the absorption towards the \ha~line assuming a \cite{Calzetti1999} dust law in the bottom row of Figure~\ref{fig:BDec}. The width ($0.2^{\prime\prime}$) and length ($0.5^{\prime\prime}$) of a typical {\it JWST} NIRSpec shutter at $z=1.7$ are shown as the dark and light grey shaded regions in Figure~\ref{fig:BDec}.

\section{Results \& Discussion}
\label{sec:discussion}


Regardless of redshift, emission-line galaxies with Log$(M_{*}/\mathrm{M_{\odot}})<9$ have centrally concentrated profiles within 1 kpc, reaching a maximum of $A_{\mathrm{H}\alpha} = 1.03\pm0.04$ mag in our high redshift sample within $ 0.4$~kpc from their center. Emission-line galaxies with $9.0\leqslant$Log($M_{*}$/M$_{\odot}$)$<10.0$ regardless of redshift have negative dust attenuation profiles. The high and low redshift samples reach maximums of $1.07\pm0.04$ and $0.79\pm0.03$ mag within $0.4$~kpc from their centers and are consistent with no dust attenuation at 3 kpc. In our highest mass bin, we measure flat dust attenuation profiles for both the high and low redshift samples. However, dust attenuation is higher at all radii for our high redshift sample, with a maximum of $1.6\pm0.1$ mag more dust attenuation within $0.4$~kpc at $1.7\leqslant z < 2.4$ than at $1.0\leqslant z < 1.7$.

We note that the two inner radial points for our low redshift high mass Balmer Decrement profile fall below the intrinsic Balmer Decrement expected in the case of no dust attenuation. These spurious measurements are likely due to contamination subtraction issues, where there is under- or over-subtraction in different locations, particularly hampering measurements in regions of low SNR, as is also evident in the second radial points of Balmer Decrement profiles for the lowest mass galaxies. High mass galaxies have the highest levels of dust attenuation (see e.g. Figure~\ref{fig:int_BDs}) with low \hb~SNRs and the lowest mass galaxies have the lowest \ha~fluxes (e.g. Figure~\ref{fig:sfms}). A combination of significantly increasing our sample of galaxies in these two stellar mass bins and actively improving our data processing algorithms (see e.g. \citealt{Matharu2022b}) will improve such measurements in future work.

\subsection{Literature Comparison}

In the two panels on the bottom right of Figure~\ref{fig:BDec}, we show results from previous works that have measured A$_{\mathrm{H}\alpha}$ in high redshift galaxies assuming a \cite{Calzetti1999} dust law. \cite{Nelson2016a}  used {\it HST} WFC3 slitless spectroscopy to measure spatially resolved Balmer Decrements at $z\sim1.4$ in 3D-HST with clean \ha~and~\hb~emission line maps and F140W magnitude $\leqslant 24$. Their results are shown as the light blue dashed and dotted lines with error envelopes and overlap in redshift with our low redshift sample (dark blue lines). For the  $9.0\leqslant$Log($M_{*}$/M$_{\odot}$)$<10.0$ stellar mass bin, the median stellar mass of our low redshift sample falls within the stellar mass range of the $9.2\leqslant$Log($M_{*}$/M$_{\odot}$)$<9.8$ (see Figure~\ref{fig:int_BDs}) $z\sim1.4$ dust attenuation profile from \cite{Nelson2016a}. Our dust attenuation profile is consistent with the \cite{Nelson2016a} dust attenuation profile within the $1\upsigma$ uncertainties of our measurements. In the higher mass bin, we do not measure a highly centrally concentrated negative dust attenuation profile like in \cite{Nelson2016a}, but a flat profile that is consistent with the \cite{Nelson2016a} profile within $1\upsigma$ of our uncertainties at radii $>2$~kpc.

The brown dotted line with solid lines delineating the $1\upsigma$ scatter shows the A$_{\mathrm{H}\alpha}$ profile inferred from the UV attenuation profile, A$_{\mathrm{UV}}$, for 10 massive star-forming main sequence galaxies at $z\sim2$ from \cite{Tacchella2018}. They recover a moderately negative A$_{\mathrm{H}\alpha}$ profile compared to our flat profile (red line). Our dust attenuation profile is consistent with the \cite{Tacchella2018} beyond 2 kpc within $1\upsigma$ of our uncertainties, but lies outside of their $1\upsigma$ scatter below these galactocentric radii. 

The disagreement in our measurements at galactocentric radii $<2$~kpc with \cite{Nelson2016a} and \cite{Tacchella2018} for our high mass bin could be due to the challenges in data processing we discussed earlier in this section and are actively working on. They could also be attributed to the large variety of galaxy-to-galaxy dust attenuation profiles measured in the literature (see e.g. \citealt{Tacchella2018}) and our small sample size in this stellar mass bin being unable to capture the true average profile of the parent distribution. Our methodology closely follows that of \cite{Nelson2016a}, but the comparison to these two works are not direct, since it can be seen in Figure~\ref{fig:BDec} that our stellar mass ranges do not match. As larger samples of spatially resolved \ha~and~\hb~emission line maps become available from {\it JWST} NIRISS and NIRCam slitless spectroscopy, we will be able to capture a larger variety of \ha~emitting galaxies and provide a more complete picture of spatially resolved dust attenuation towards \ion{H}{2} regions out to $z\sim5$.

\subsection{How Dust Gradients effect SFR measurements from Slit Spectroscopy}
\label{sec:sfr_cons}
We test the effect of assuming uniform dust attenuation versus the dust attenuation profiles we measure, to SFR measurements of typical star-forming galaxies at these redshifts and stellar masses. We create PSF-convolved \texttt{GALFIT} models for each of our redshift and stellar mass bins with the half-light radii and S\'ersic indices typical of star-forming galaxies \citep{Mowla2019} but with the brightnesses of our \ha~stacks. We then dust-correct our models using a single value of $\mathrm{A}_{\mathrm{H}\alpha}$ calculated from the integrated Balmer Decrements in Figure~\ref{fig:int_BDs} and with a 2D map of our $\mathrm{A}_{\mathrm{H}\alpha}$ radial profiles shown in Figure~\ref{fig:BDec}. On average, we find that assuming uniform dust attenuation across the galaxy subsequent to slit loss corrections will overestimate the SFRs of typical star-forming galaxies by $6\pm21 \%$ and $26\pm9 \%$  of the SFR calculated with the radial dust attenuation corrections based on our profiles at $1.0\leqslant z < 1.7$ and $1.7\leqslant z < 2.4$ respectively.

\

In summary, our measurements indicate that:

\begin{enumerate}

    \item Emission-line galaxies with $7.6\leqslant$Log($M_{*}$/M$_{\odot}$)$<10.0$ have centrally concentrated, negative dust attenuation profiles. At the lowest masses ($7.6\leqslant$Log($M_{*}$/M$_{\odot}$)$<9.0$), maxmimum attenuations of $A_{\mathrm{H}\alpha} = 0.93\pm0.05$ and $1.03\pm0.04$ mag are reached at $1.0\leqslant z < 1.7$ and $1.7\leqslant z < 2.4$ within $0.4$~kpc from the center respectively. At $9.0\leqslant$Log($M_{*}$/M$_{\odot}$)$<10.0$, these values are $0.79\pm0.03$ and $1.07\pm0.04$ mag respectively.

    \item Emission-line galaxies with $10.0\leqslant$Log($M_{*}$/M$_{\odot}$)$<11.1$ have flat dust attenuation profiles with an average of $1.0\pm0.2$ mag more dust attenuation measured at $1.7\leqslant z < 2.4$ than at $1.0\leqslant z < 1.7$.

    \item Assuming uniform dust attenuation rather than the radial profiles we measure subsequent to (or during) slit loss corrections will overestimate \ha~SFR measurements by an average of $6\pm21\%$ and $26\pm9\%$ at $1.0\leqslant z < 1.7$ and $1.7\leqslant z < 2.4$ respectively (Section~\ref{sec:sfr_cons}).

\end{enumerate}

\acknowledgments

\noindent JM is grateful to the Cosmic Dawn Center for the DAWN Fellowship. JM would like to thank Ivelina Momcheva, Erica Nelson,  Taylor Hutchison and Conor McPartland for useful discussions that led to improvements in the analysis presented in this paper. This research was enabled by grant 18JWST-GTO1 from the Canadian Space Agency and funding from the Natural Sciences and Engineering Research Council of Canada. MB acknowledges support from the Slovenian national research agency ARRS through grant N1-0238 and the program HST-GO-16667, provided through a grant from the STScI under NASA contract NAS5-26555. This research used the Canadian Advanced Network For Astronomy Research (CANFAR) operated in partnership by the Canadian Astronomy Data Centre and The Digital Research Alliance of Canada with support from the National Research Council of Canada the Canadian Space Agency, CANARIE and the Canadian Foundation for Innovation.

\software{This research made use of \textsc{Astropy}, a community-developed core Python package for Astronomy \citep{TheAstropyCollaboration2018}. The python packages \textsc{Matplotlib} \citep{Hunter2007}, \textsc{Numpy} \citep{VanDerWalt2011} and \textsc{Scipy} \citep{scipy} were also extensively used. Parts of the results in this work make use of the colormaps in the \textsc{CMasher} \citep{VanderVelden2020} package. Parts of the results in this work use color palettes from \textsc{Palettable}\footnote{\url{https://jiffyclub.github.io/palettable/}}.}

\facilities{ {\it JWST}\ (NASA/ESA/CSA), {\it HST}\ (NASA/ESA)}


\bibliography{library_grizli}{}
\bibliographystyle{aasjournal}



\end{document}